# Unveiling Orbital-mediated Ultrafast Demagnetization in Rare-Earth-Transition-Metal Ferrimagnets

Jianwen Gao[1], Linlin Zhang[1], Mingli Ge[1], Runhua Zhang[1], Jinshan Wang[1], Hui Li[1], Xiaowei Zhou[1], Zhu Liu[1], Zongzhi Zhang[2]*, Li Xi[3]*,Yalu Zuo[3], Chenglong Jia[3], Feng Qiu[4], Shaojie Hu[5], Yang Ren[1]*

[1]*School of Physics and Astronomy, Yunnan University, Kunming 650091, China.*

[2]*Key Laboratory of Micro and Nano Photonic Structures (MOE), College of Future Information Technology, Fudan University, Shanghai 200433, China.*

[3]*Key Laboratory for Magnetism and Magnetic Materials of Ministry of Education, School of Physical Science and Technology, Lanzhou University, Lanzhou 730000, China.*

[4]*School of Materials and Energy, Yunnan University, Kunming 650091, China.*

[5]*Engineering Research Center of Guangdong for Compound Semiconductor Devices and Chips, College of Integrated Circuits and Optoelectronic Chips, Shenzhen Technology University, 3002 Lantian Road, Pingshan District, Shenzhen Guangdong 518118, China.*

## Abstract

The ultimate speed limit of magnetic recording and spintronic devices is set by the efficiency of angular-momentum transfer during ultrafast demagnetization, yet its microscopic pathway in Rare-Earth-Transition-Metal (RE-TM) ferrimagnets remains debated. Here, we establish an orbital-mediated framework in which 3*d* spin-orbit coupling (SOC) governs angular momentum (AM) dissipation. Strong 3*d*-SOC in RE-Co enables sub-picosecond, single-step demagnetization via direct orbital-to-lattice transfer, whereas weak 3*d*-SOC in RE-Fe redirects AM into 4*f* orbitals, producing slower two-step dynamics. The second-stage rate scales with 4*f*-SOC strength, revealing a distinct orbital-mediated dissipation channel. Using time-resolved magneto-optical Kerr measurements, supported by an extended four-temperature model, corroborate this picture across diverse RE-TM systems (RE = Sm, Gd, Tb, Dy, Ho and TM = Fe, Co, CoNi). Our results identify the SOC-driven competition between 3*d* and 4*f* orbital channels as the universal mechanism governing ultrafast demagnetization in RE-TM ferrimagnets, enabling rational design of the switching speed for next-generation spintronic devices.

*Corresponding authors: reny@ynu.edu.cn (Yang Ren), zzzhang@fudan.edu.cn (Zongzhi Zhang), xili@lzu.edu.cn (Li Xi) .

The ultimate speed limit of magnetic recording and spintronic devices is determined by the efficiency of angular momentum (AM) transfer during ultrafast magnetization dynamics [1-3]. In simple transition-metal (TM) ferromagnets such as Ni, Fe, and Co, ultrafast demagnetization occurs on a well-defined sub-picosecond timescale, driven predominantly by spin-orbit coupling (SOC)-mediated spin-flip scattering [4-6]. Rare-Earth-Transition-Metal (RE-TM) ferrimagnets, in stark contrast, exhibit a dramatically wider range of demagnetization times, from sub-picoseconds to tens of picoseconds, depending on the specific RE and TM constituents [7-9]. This striking diversity, together with their antiferromagnetic sublattice coupling and tunable compensation points, renders RE-TM materials essential for ultrafast magnetic manipulation and applications, including all-optical switching, picosecond spin-orbit torque switching, terahertz emission, ultrafast domain-wall motion and et al [10-17]. Yet, despite this broad interest, the microscopic angular-momentum transfer mechanism responsible for the remarkable diversity in RE-TM demagnetization dynamics remains unclear [8, 18, 19].

Ultrafast demagnetization in TMs is well described by the Elliott-Yafet (EY) model, in which SOC mediates electron-phonon spin-flip scattering that transfers AM to the lattice [5, 20, 21]. Koopmans et al. extended this framework to RE metals [22]. In some RE-TM ferrimagnets, however, intersublattice exchange coupling dominates the ultrafast dynamics: Radu et al. revealed that spin reversal in GdFeCo proceeds via a transient ferromagnetic-like state mediated by strong $3d$-$4f$ exchange [23], and Mentink et al. further established this exchange as an additional heat transport channel [8].

While intersublattice exchange clearly mediates the transient ferromagnetic-like state in RE-TM alloys, element-specific X-ray Magnetic Circular Dichroism (XMCD) measurements have separately established that SOC actively governs the ultrafast spin dynamics of both $3d$ and $4f$ spins, respectively [24-27]. Fe in GdFeCo exhibits a strictly constant orbital-to-spin ratio during demagnetization [24], whereas Co in CoGd and CoTb undergoes a pronounced transient quenching of this ratio [25]. In contrast, Tb carries a large orbital moment, whereas Gd exhibits no orbital moment [25]. Crucially, Hennecke et al. [24] observed complete transfer of spin and orbital AM to the lattice within hundreds of femtoseconds in GdFeCo. Nevertheless, how 3*d*-SOC, 4*f*-SOC, and intersublattice 3*d*-4*f* exchange together govern the AM transfer remains unexplored.

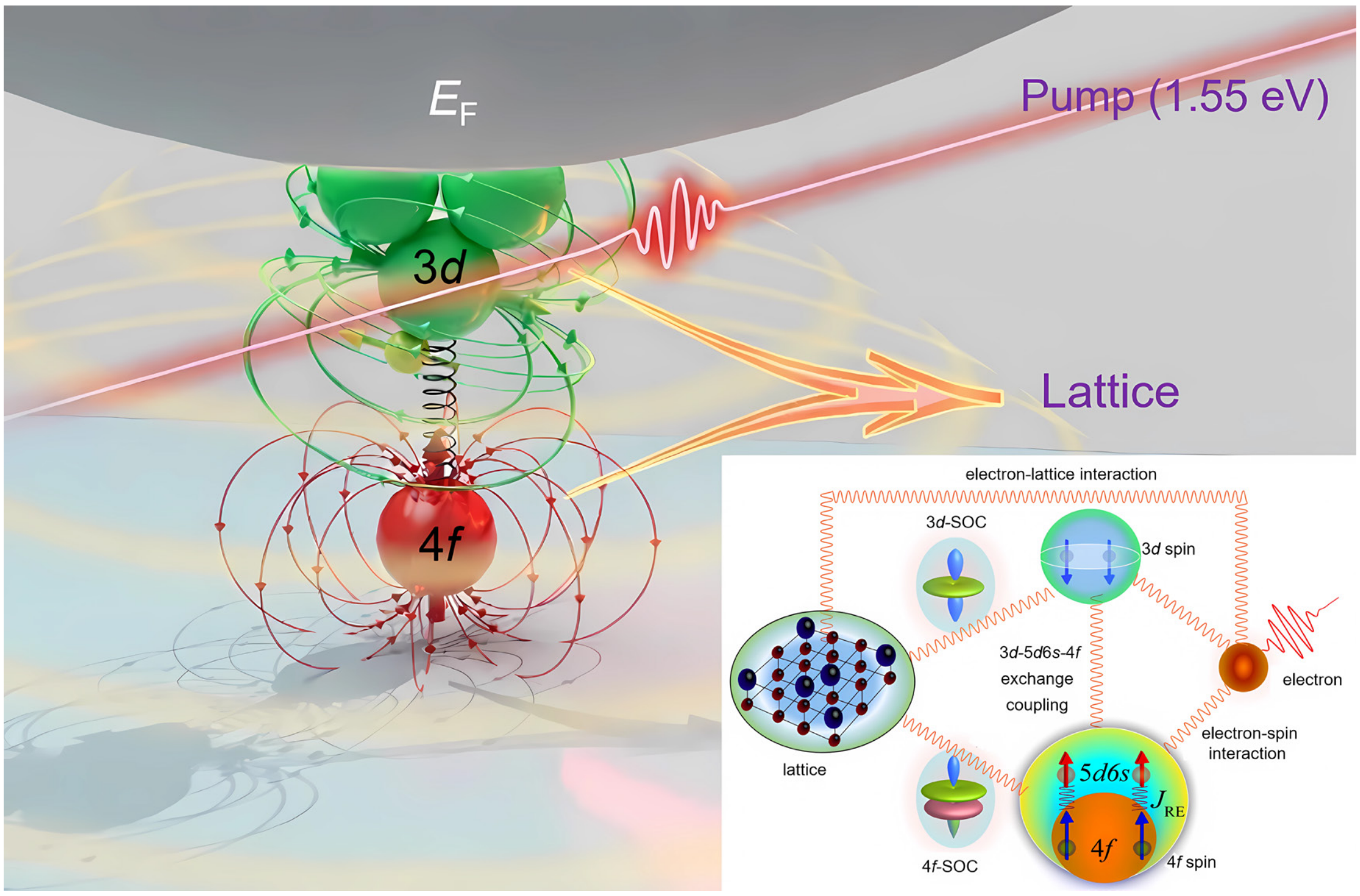

**Fig.1 | AM dissipation pathways in RE-TM alloys.** Schematic illustration of the proposed SOC-mediated mechanism for AM transfer in RE-TM alloys following fs-infrared laser excitation. The redistribution pathways are governed by the interplay of 3*d* and 4*f* SOC and intersublattice exchange, leading to distinct demagnetization channels.

Here we establish an orbital-mediated framework that resolves this bottleneck, illustrated in Fig. 1. Upon femtosecond laser excitation, the redistribution of energy and AM is governed by the competition between three key ingredients: the 3*d*-SOC strength, the 4*f*-SOC strength, and the intersublattice exchange coupling. When the 3*d*-SOC is strong, energy and AM is rapidly transferred from spin to orbital degrees of freedom and then efficiently dissipated into the lattice, yielding single-step sub-picosecond demagnetization. When the 3*d*-SOC is weak, the direct TM orbital-to-lattice channel is suppressed; instead, a fraction of energy and AM is funneled into RE 4*f* orbitals via 3*d*-5*d*6*s*-4*f* pathways, resulting in slower, multistep dynamics. The role of the RE element is further dictated by its orbital character: systems with orbital quantum number $Q_L \neq 0$ support both 4*f* orbital-mediated dissipation to the lattice and back-transfer to the 3*d* subsystem, while $Q_L=0$ suppresses 4*f* orbital dissipation, leading to dominant backflow and slower demagnetization. This scenario establishes a unified framework for demagnetization in RE-TM ferrimagnets and identifies the relative strength of 3*d* SOC versus 4*f* SOC as a decisive control parameter for ultrafast energy and AM pathways.

To validate this picture, we perform systematic time-resolved magneto-optical Kerr effect (TR-MOKE) measurements on RE-TM alloys (RE=Sm, Gd, Tb, Dy, Ho with various $Q_L$; TM=Fe, Co, CoNi). The 3*d*-SOC strength increases from Fe (≈70 meV) to Co (≈90 meV) to Ni (≈110 meV) [28]. We find that RE-Co exhibits single-step demagnetization consistent with strong 3*d* SOC-driven orbital dissipation, whereas RE-Fe shows a characteristic two-step dynamic, with the slower stage governed by intersublattice coupling and 4*f* orbital channels. Controlled Ni doping enhances 3*d* SOC strength and systematically accelerates demagnetization, providing more evidence for the proposed mechanism. The experimental results are in excellent agreement with the theoretical calculation. Our results identify the competition between 3*d*- and 4*f*- SOC as the key determinant of ultrafast energy and AM transfer and establish the 3*d* SOC strength as its dominant tuning parameter in RE-TM ferrimagnets.

Figures 2a and 2b present the ultrafast demagnetization dynamics of $RE_{0.15}TM_{0.85}$ alloy thin films (RE = Sm, Gd, Tb, Dy, Ho; TM = Co, Fe), measured using TR-MOKE spectroscopy at a pump fluence of 1.5 $mJ/cm^2$ under a saturating magnetic field of 6 kOe. Two distinct dynamical regimes are clearly resolved. RE-Co alloys exhibit a single-step sub-ps demagnetization followed by rapid recovery, closely resembling the response of pure 3*d* transition metals [1, 29]. In contrast, the RE-Fe alloys display two-step dynamics, consisting of an initial rapid drop within a few ps followed by a slower decay extending to hundreds of ps, reminiscent of behavior reported in Gd/Tb films [30].

This contrast directly reflects the role of TM spin-orbit coupling. Control measurements on pure Co and Fe (Fig. S2 in SI) both show single-step sub-ps demagnetization, indicating that the emergence of two-step dynamics in RE-Fe originates from the coupling to RE degrees of freedom rather than an intrinsic property of Fe. In the strong SOC case (Co), energy and AM is efficiently transferred to the lattice, leading to single-step demagnetization. In the weaker SOC case (Fe), apart from the direct transfer from spin to lattice, partial energy and AM is redirected into RE 4*f* orbitals via 3*d*-5*d*6*s*-4*f* channels, giving rise to a secondary, slower relaxation process.

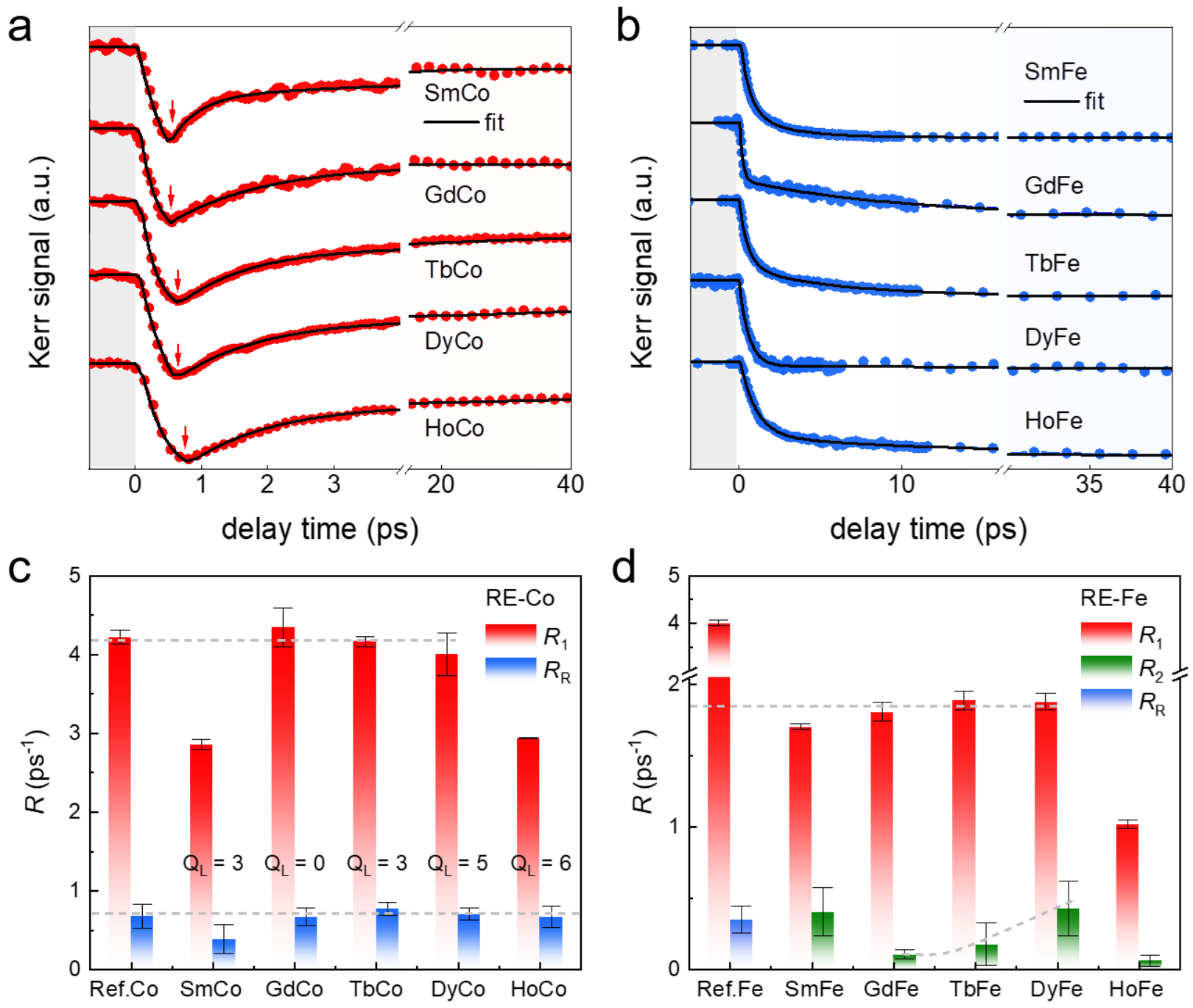

**Fig.2 | Ultrafast demagnetization dynamics of RE-TM systems.** Normalized TR-MOKE signals of ultrafast demagnetization for a, $RE_{0.15}Co_{0.85}$ (RE-Co) and b, $RE_{0.15}Fe_{0.85}$ (RE-Fe) films. Red arrows indicate shifts in signal extrema, and solid black lines represent biexponential fits. c, Extracted first-step demagnetization rate $R_1$ and recovery rate $R_R$ for RE-Co alloys, with pure Co as a reference. d, Extracted demagnetization and recovery rates $R_1$, $R_2$, and $R_R$ for RE-Fe alloys, with pure Fe as a reference.

To further disentangle the role of TM SOC from RE composition, we systematically varied the composition in $RE_xTM_{1-x}$ films (see Supplementary Note 2). The qualitative behavior remains unchanged across all compositions, with single-step dynamics in RE-Co and two-step dynamics in RE-Fe, demonstrating that the demagnetization pathway is governed primarily by the TM SOC rather than RE concentration.

This trend is robust over a wide composition range, with one notable exception: $Gd_{0.30}Fe_{0.70}$ exhibits an anomalous discontinuous two-step demagnetization process (Fig. S3 in SI), consistent with prior reports [8, 31]. This deviation originates from the vanishing orbital moment in Gd, which blocks the efficient 4*f*-SOC-mediated momentum transfer to lattice. As a result, energy dissipation is forced to proceed through weaker intersublattice exchange pathways involving 3*d*-5*d*6*s*-4*f*, leading to a delayed and discontinuous two-step response [31].

By fitting the TR-MOKE data with a bi-exponential function [7, 18] (see Supplementary Note 3), we extract three characteristic timescales: the first-step demagnetization time $\tau_1$, the second-step demagnetization time $\tau_2$, and the magnetization recovery time $\tau_R$. The corresponding rates, $R_1 = 1/\tau_1$, $R_2 = 1/\tau_2$, and $R_R = 1/\tau_R$, are plotted in Figs. 2c and 2d. We find that $R_1$ in RE-Co alloys is nearly twice that in their RE-Fe counterparts, indicating a more efficient energy and AM transfer enabled by the stronger 3*d*-SOC of Co [28, 32]. For most RE-Co alloys (RE = Gd, Tb, Dy) as well as pure Co, $R_1$ and $R_R$ are comparable, suggesting that energy dissipation is predominantly governed by 3*d*-orbital-mediated processes, driven by strong 3*d* SOC.

In contrast, SmCo and HoCo exhibit reduced $R_1$, which we attribute to their weak room-temperature magnetization: the paramagnetic nature of Sm and Ho weakens the Co-Co exchange coupling [33, 34], thereby slowing the initial demagnetization. Notably, HoCo shows a larger $R_R$ than SmCo, indicating faster recovery, which can be ascribed to the larger orbital AM of Ho ($Q_L = 6$) that enhances 4*f*-SOC-mediated orbital-lattice coupling.

For RE-Fe alloys (RE=Gd, Tb, Dy), $R_1$ is nearly RE-independent yet about a factor of 2 smaller than in pure Fe, indicating a secondary role of RE elements in the initial demagnetization. In contrast, $R_2$ increases monotonically with the $Q_L$ of RE. This trend can be understood in terms of the relatively weak SOC of Fe, which leads to two competing energy and AM dissipation channels: a direct Fe 3*d* SOC and a slower pathway mediated by 3*d*-5*d*6*s*-4*f* interactions. The latter becomes increasingly efficient with larger 4*f* orbital moments, giving rise to the observed enhancement of $R_2$.

The enhancement of $R_2$ confirms the role of 4*f*-SOC in facilitating spin-lattice energy transfer [18]. In GdFe, the weak Gd-SOC leads to energy backflow from the 4*f* to the 3*d* subsystem, reheating the spin system and slowing dissipation [31]. By contrast, for TbFe and DyFe, energy transferred to the 4*f* subsystem splits into two channels: backflow to the 3*d* spins and direct dissipation into the lattice via strong 4*f*-SOC-mediated orbital-lattice coupling, yielding faster $R_2$.

HoFe, characterized by low Curie temperatures, exhibit suppressed $R_1$ and $R_2$, which can be attributed to inefficient energy and AM transfer arising from weakened magnetic exchange [33, 34]. SmFe exhibit reduced $R_1$ due to its weak room-temperature magnetization. However, SmFe (the light RE-TM alloy) exhibits weak antiferromagnetic coupling, thus only a small fraction of AM is transferred from the 3*d* spins to the 4*f* spins, enabling faster dissipation via 4*f* SOC, resulting in a faster $R_2$. Composition-dependent measurements (see Supplementary Note 2) further reveal that $R_1$ is largely insensitive to RE concentration and is primarily governed by the 3*d*-SOC strength. In contrast, $R_2$ exhibits a more complex dependence, governed by the interplay of 3*d*-SOC, 4*f*-SOC, intersublattice 3*d*-4*f* coupling, and the intrinsic magnetic properties of the RE-TM alloys.

The distinct demagnetization dynamics, single-step in RE-Co and two-step in RE-Fe, identify the TM's SOC strength as the key control parameter. This trend implies that further increasing the 3*d* SOC strength should continuously reshape the demagnetization landscape. Motivated by this rationale, we turn to nickel (Ni), which has a stronger SOC than Co and is thus expected to enhance energy and AM dissipation [35]. Substituting Ni into RE-Co amorphous films should therefore preserve the single-step characteristic [7] while substantially accelerating the demagnetization rate. However, before proceeding, it is worth clarifying that direct RE-Ni alloys are unsuitable for room-temperature ultrafast spintronics. The Curie temperatures of RE-Ni systems lie well below room temperature [36], and femtosecond laser excitation risks efficient spin quenching in Ni-rich systems [37]. Moreover, RE-Ni thin films suffer from poor compositional and structural stability [38]. To circumvent these limitations while still harnessing Ni's strong SOC, we introduced controlled amounts of Ni into TbCo and GdCo alloys (demagnetization curves for TbCoNi and GdCoNi are shown in Fig. 3a and Supplementary Fig S8a, respectively). As shown in Fig. 3a, TbCoNi alloys exhibit a sub-ps, single-step demagnetization followed by rapid recovery, confirming the dominant role of the strong SOC in CoNi matrix. Biexponential fits quantify the increase of $R_1$ with Ni concentration (Fig. 3a, inset), consistent with prior observations that enhanced $3d$-SOC improves energy dissipation. This behavior is mirrored in GdCoNi alloys, where $R_1$ also exhibits a clear rise with increasing Ni content (Supplementary Fig. S8a). These findings further confirm that strong $3d$-SOC in RE-TM systems enables rapid dissipation of energy and AM from the $3d$-orbitals to the lattice, thereby accelerating the demagnetization and yielding single-step dynamics.

To further elucidate ultrafast demagnetization in RE-TM alloys, we developed an extended four-temperature model (E4TM, see Supplementary Note 3), explicitly incorporating SOC and inter-sublattice $3d$-$4f$ exchange [8, 9]. The model tracks energy exchange among electrons ($T_e$), lattice ($T_l$), TM $3d$ spins ($T_s^{TM}$), and RE $4f$ spins ($T_s^{RE}$). The SOC-mediated energy dissipation is characterized by two material-specific constants: $G_{soc}^{3d}$ and $G_{soc}^{4f}$, which represent the spin-to-lattice energy transfer efficiencies and correspondingly reflect the strengths of $3d$-SOC and $4f$-SOC in the TM and RE subsystems, respectively.

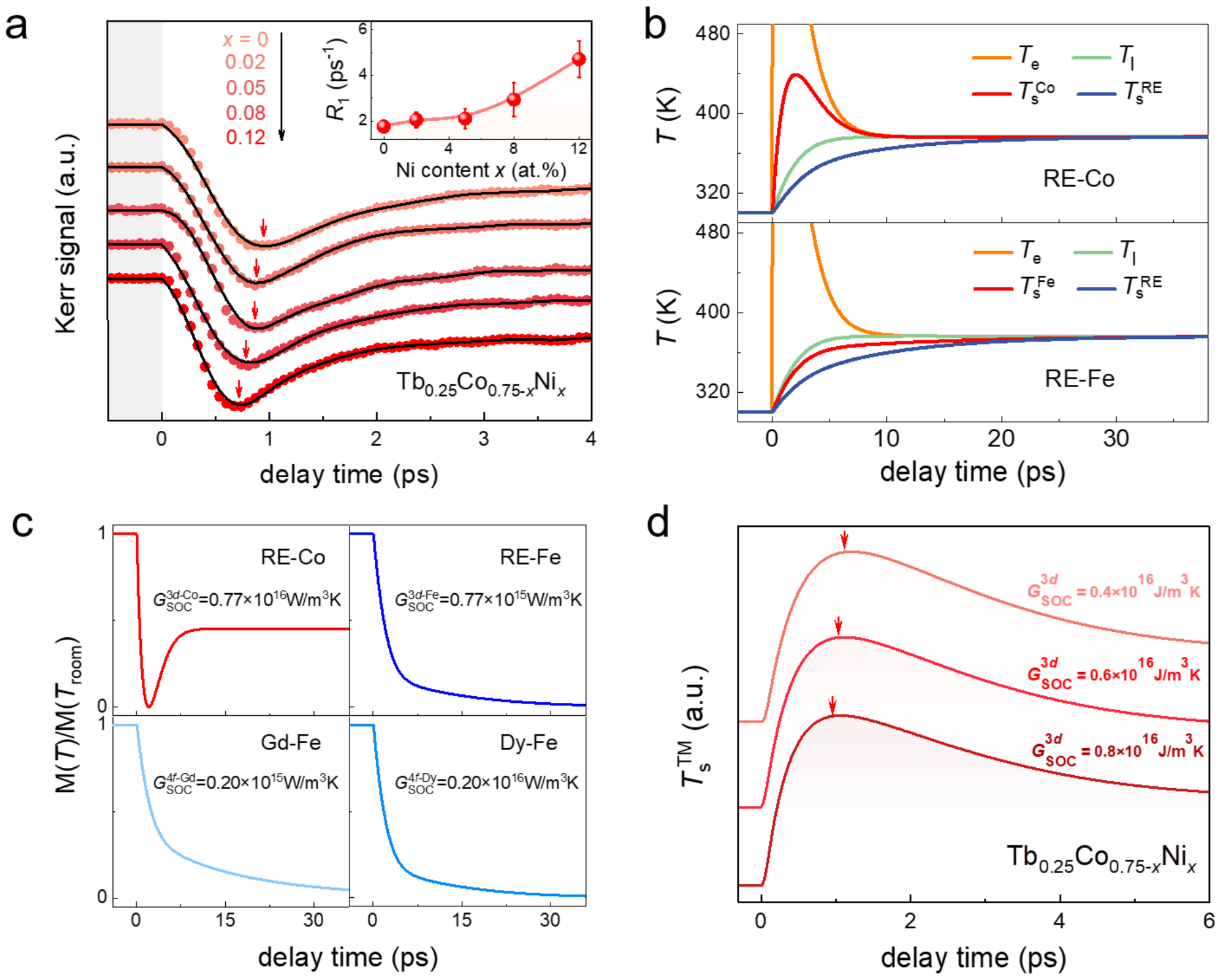

**Fig.3 | Ultrafast demagnetization dynamics of Ni-doped RE-Co systems and E4TM analysis in RE-TM systems.** a, TR-MOKE traces of Ni-doped $Tb_{0.25}Co_{0.75}$ films. Red arrows indicate shifts in signal extrema, and solid black lines represent biexponential fits. The inset displays the initial demagnetization rate $R_1$ as a function of Ni content. b, Temporal evolution of the heat-reservoir temperatures ($T_e$, $T_l$, $T_s^{TM}$, and $T_s^{RE}$) in RE-Co and RE-Fe alloys. c, Magnetization dynamics in RE-TM alloys with varying $G_{SOC}$, derived using the Bloch $T^{3/2}$ law. d, Simulated temporal evolution of $T_s^{TM}$ for $Tb_{0.25}Co_{0.75\text{-}x}Ni_x$ alloys.

In typical RE-Co alloys (Fig. 3b, upper panel), the large $G_{\mathrm{soc}}^{3d\text{-Co}}$ ($0.77 \times 10^{16}$ W/m$^3$K) drives $T_s^{\mathrm{Co}}$ to rise rapidly to approximately 440 K within 2 ps, after which it equilibrates with the $T_e$ and $T_l$ temperatures within 10 ps, demonstrating highly efficient orbital-mediated spin-lattice energy transfer. In contrast, RE-Fe alloys (Fig. 3b, lower panel) exhibit a much smaller $G_{\mathrm{soc}}^{3d\text{-Fe}}$ ($0.77 \times 10^{15}$ W/m$^3$K). Consequently, the $T_s^{\mathrm{Fe}}$ rises only to ~350 K at 2.8 ps, and the equilibration between $T_e$ and $T_l$ is not achieved until after 20 ps. Such behavior indicates that a small SOC strength limits the 3$d$ orbital-mediated pathway, forcing the system to rely on slower, less efficient channels. The low $G_{\mathrm{soc}}^{3d\text{-Fe}}$ activates a distinct pathway: energy initially flows from Fe-3$d$ to RE-4$f$ spins via 3$d$-4$f$ exchange interaction, then partially dissipates via RE-SOC-mediated orbital-lattice interactions and partially backflows to the 3$d$ subsystem. This bidirectional energy flow between 3$d$/4$f$ spins and lattice gives rise to the two-step demagnetization observed experimentally. More Detailed description of the calculation results is provided in the Supplementary Note 3. Converting the simulated $T_s^{TM}$ into magnetization using the Bloch $T^{3/2}$ law [8], $M(T)/M(T_{room}) = 1 - (T/T_C)^{3/2}$, yields magnetization curves (Fig. 3c) that accurately reproduce the TR-MOKE data. These results validate that the E4TM model captures the essential physics of RE-TM demagnetization by incorporating both SOC and inter-sublattice exchange coupling.

Beyond the dominant role of 3$d$-SOC in RE-Co systems, the contribution of 4$f$-SOC becomes particularly significant in RE-Fe alloys. Comparative simulations of GdFe ($Q_L$= 0 for Gd) and DyFe ($Q_L$= 5 for Dy) illustrate this distinction (see Fig. 3c; see also Fig. S7 in the SI for details). By setting $G_{soc}^{4f\text{-}Gd} = 0.20 \times 10^{15}$ W/m$^3$K and $G_{soc}^{4f\text{-}Dy} = 2.0 \times 10^{15}$ W/m$^3$K, we observe that the temporal evolution of $T_s^{Fe}$ depends strongly on the RE species. Compared to the weak SOC of Gd, Dy's strong 4$f$-SOC accelerates the demagnetization dynamics.

The above analysis of the E4TM successfully reproduces the experimentally observed distinction between RE-Co and RE-Fe systems, confirming that the 3$d$-/4$f$-SOC governs the efficiency of energy and AM transfer. Applying the theoretical calculation to the $Tb_{0.25}Co_{0.75-x}Ni_x$ system, the simulated magnetization curves capture the experimental trend well: as $G_{soc}^{3d}$ increases from $0.4\times10^{16}$ to $0.8\times10^{16}$ J/m$^3$K (Fig.3d), the simulated dynamics closely match the experimental results. A similar behavior is also observed in GdCoNi alloys (see Supplementary Note 4). Together, these results demonstrate that 3$d$-SOC engineering serves as a highly effective approach for tailoring spin dynamics in RE-TM materials.

Finally, terahertz (THz) emission measurements provide independent confirmation of enhanced energy dissipation pathways in Gd-based RE alloys (see Fig.

S9 in the SI). The THz intensity, which is proportional to the laser-induced orbital current ($J_L$), is notably higher in GdCo than in GdFe, indicating that the stronger 3*d*-SOC of Co generates orbital angular momentum, which is then converted into charge current via the inverse orbital Hall effect, thereby enhanced the THz emission efficiency. This observation aligns with established findings that stronger spin-orbit coupling (SOC) generates a more efficient orbital current, thereby enhancing THz emission intensity [16, 39].

In summary, we have established a unified orbital-mediated framework that resolves the controversy over ultrafast demagnetization in RE-TM ferrimagnets. The key is a competition hierarchy: the 3*d*-SOC strength determines whether energy and AM are efficiently funneled into orbital-to-lattice dissipation (strong 3*d*-SOC, single-step) or redirected into 4*f* orbital channels (weak SOC, two-step). This competition is further modulated by the RE orbital character: $Q \neq 0$ (e.g., Tb) enables both 4*f*-SOC-mediated lattice dissipation and backflow, whereas $Q_L = 0$ (e.g., Gd) blocks orbital dissipation, forcing energy and AM to return to the 3*d* sublattice. Systematic Ni doping, which continuously enhances the effective 3*d* SOC strength, provides quantitative confirmation that 3*d* SOC, rather than intersublattice exchange alone, dictates the dominant dissipation route. By capturing these intricate processes within an extended four-temperature model analysis, we demonstrate that our orbital-mediated framework thus offers a universal account of RE-TM demagnetization phenomena and a predictive guide for the rational design of ultrafast spintronic devices.

## Methods

Magnetic multilayer samples were prepared on $Si/SiO_2$ substrate by DC magnetron sputtering at room temperature. The multilayers consist of Ta (5 nm)/RE-TM (20 nm) /Ta (3 nm)/substrate, where the RE represents one of the RE metals: Sm, Tb, Gd, Dy or Ho and TM represents Fe, Co or Co-Ni. The base pressure was below $5\times10^{-8}$ Torr and a working Ar pressure of 5.0 mTorr was kept unchanged during the sputtering. The composition of the RE-TM alloys was measured by energy dispersive X-Ray spectroscopy. Magnetic properties were checked by magneto-optical Kerr effect (MOKE) technique with both longitudinal and polar geometry.

A Time resolved Magneto-optical Kerr spectroscopy (TR-MOKE) with both longitudinal and polar geometry were employed for the detection of ultrafast demagnetization process. The system was driven by a Ti: sapphire laser at a wavelength of 800 nm, with a repetition rate of 1 kHz and a pulse width of about 100 fs. After splitting the beam into two linearly polarized parts with unequal power, the stronger pump pulse incident at 10° from the sample normal was focused to a spot diameter of around 0.5 mm, while the weaker probe pulse incident at 45° from the sample normal, controllably delayed, was focused to a smaller spot diameter which was around 50 μm. The larger spot size of the pump over the probe beam ensures homogeneous heating in the probed area of the sample. The magnetization was locked along its easy direction by an external magnetic field, the magnitude of which was greater than the coercive field. The present experimental geometries at room temperature avoid laser-induced magnetization precessions as well as laser-induced switching across a ferrimagnetic compensation point.

## Supplemental Information

Supplemental Information is available in the online version of the paper or from the author.

## Acknowledgements

This work was primarily supported by the National Natural Science Foundation of China (Grant Nos. 12464016, 12164053, 12264054, 12574120), Double First Class Joint Special Key Project of Yunnan Science and Technology Department & Yunnan University (Grant No. 202401BF070001-012), Shenzhen Science and Technology Program (JCYJ20240813113228037).

## Data Availability Statement

The data that support the findings of this study are available from the corresponding author upon reasonable request.

# Supplemental Information: Unveiling Orbital-mediated Ultrafast Demagnetization in Rare-Earth-Transition-Metal Ferrimagnets

Jianwen Gao[1], Linlin Zhang[1], Mingli Ge[1], Runhua Zhang[1], Jinshan Wang[1], Hui Li[1], Xiaowei Zhou[1], Zhu Liu[1], Zongzhi Zhang[2]*, Li Xi[3]*,Yalu Zuo[3], Chenglong Jia[3], Feng Qiu[4], Shaojie Hu[5], Yang Ren[1]*

[1]*School of Physics and Astronomy, Yunnan University, Kunming 650091, China.*

[2]*Key Laboratory of Micro and Nano Photonic Structures (MOE), College of Future Information Technology, Fudan University, Shanghai 200433, China.*

[3]*Key Laboratory for Magnetism and Magnetic Materials of Ministry of Education, School of Physical Science and Technology, Lanzhou University, Lanzhou 730000, China.*

[4]*School of Materials and Energy, Yunnan University, Kunming 650091, China.*

[5]*Engineering Research Center of Guangdong for Compound Semiconductor Devices and Chips, College of Integrated Circuits and Optoelectronic Chips, Shenzhen Technology University, 3002 Lantian Road, Pingshan District, Shenzhen Guangdong 518118, China.*

*Corresponding authors: reny@ynu.edu.cn (Yang Ren), zzzhang@fudan.edu.cn (Zongzhi Zhang), xili@lzu.edu.cn (Li Xi) .

## Table of contents

**Supplementary Note 1: Stoichiometric analysis of typical Rare-Earth-Transition-Metal (RE-TM) alloys.**

X-ray spectroscopy (EDX) measurements were performed to verify the chemical composition of typical RE-Co alloy thin films. The representative EDX spectra are shown for GdCo (top row) and HoCo (bottom row) samples with varying RE content. Characteristic peaks corresponding to Co (magenta arrows), Gd (red arrows), and Ho (blue arrows) are marked. The extracted atomic concentrations are displayed as insets within each panel, which validates the accuracy of the compositional data presented in the manuscript.

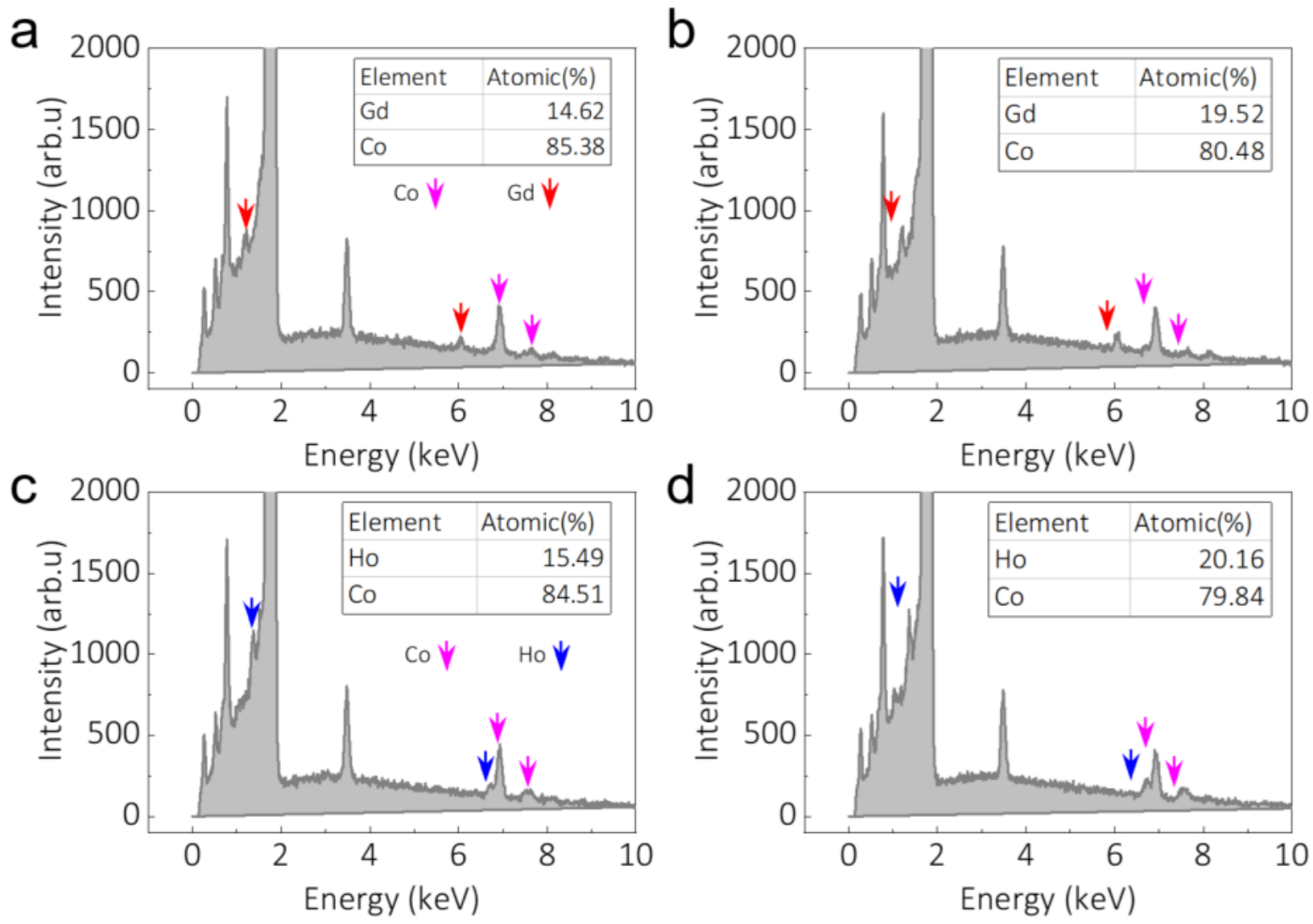

**Supplementary Figure S1.** Compositional analysis of typical RE-TM alloys by energy-dispersive X-ray spectroscopy.

**Supplementary Note 2: Ultrafast demagnetization in $RE_xTM_{1-x}$ alloys.**

To investigate the compositional dependence of demagnetization rate and isolate spin-orbit coupling (SOC) effects in rare-earth-transition-metal (RE-TM) systems, a series of $RE_xTM_{1-x}$ alloys ($x$ = 0.15 - 0.30; TM = Co or Fe) were fabricated. The RE composition range was selected to avoid regimes where the dynamics are dominated by either pure TM-like behavior ($x$ < 0.15) or significantly reduced magnetization and Curie temperature ($x > 0.30$), as reported previously [1, 2, 3]. TR-MOKE measurements performed at a pump fluence of 1.5 mJ/cm$^2$ and an external field of 6 kOe reveal clear ultrafast demagnetization in all samples (Figs. S2-S6), consistent well with theoretical predictions.

Fig. S2 compares the demagnetization dynamics of pure Co and Fe ($x$=0), both of which exhibit single-step, sub-picosecond demagnetization with nearly identical time scales ($R_1$). These observations highlight the intrinsic ultrafast nature of 3$d$ transition metals: within the first hundreds of femtoseconds (<500 fs), the TR-MOKE response is dominated by state-filling effects, whereas true demagnetization occurs on the ~0.5-1 ps [4]. Additionally, Fe exhibits a smaller magnetization recovery rate ($R_R$) than Co, indicating slower rate of orbital mediated energy transfer from the spin system to the lattice, which stems from Fe's weaker 3$d$-SOC strength.

In contrast to pure Co and Fe, $RE_xTM_{1-x}$ alloys exhibit qualitatively different demagnetization behavior. This likely stems from Co's stronger net spin magnetic moment (vs. Fe), which enhances its propensity for orbital hybridization with RE elements. A detailed investigation of this mechanism will be pursued in subsequent studies. In all $RE_xCo_{1-x}$ alloys, magnetization exhibits a rapid, sub-picosecond decay followed by fast recovery within a few picoseconds. In contrast, $RE_xFe_{1-x}$ alloys undergo an initial ultrafast decay followed by a slower component extending over hundreds of picoseconds. Notably, the single-step demagnetization rate $R_1$ in $RE_xCo_{1-x}$ alloys remain nearly constant across compositions and close to the value of pure Co, indicating that energy transfer from 3$d$ spins to the lattice is primarily governed by 3$d$-SOC via Co orbitals. Nevertheless, the recovery rates $R_R$ show composition-dependent trends: $Sm_xCo_{1-x}$ and $Gd_xCo_{1-x}$ exhibit relatively stable $R_R$, while $R_R$ in $Tb_xCo_{1-x}$, $Dy_xCo_{1-x}$, and $Ho_xCo_{1-x}$ decreases with increasing $x$, likely due to differences in structure and/or magnetic ordering [2, 5, 6].

In $RE_xFe_{1-x}$ alloys, $R_1$ is largely composition-independent, suggesting a weak Fe 3$d$-SOC channel. However, $R_2$ varies significantly with RE content, reflecting the influence of 3$d$-4$f$ intersublattice coupling and 4$f$-SOC. In $Sm_xFe_{1-x}$, $R_2$ remains nearly unchanged. This is consistent with the fact that Sm, as a light rare earth element, forms ferromagnetic alloys with Fe[7]. Consequently, Sm exhibits very weak magnetism at room temperature, and as a result, its contribution via the 4*f* channels is minimal. In $Gd_xFe_{1-x}$ alloys, both the continuous and discontinuous two-stage demagnetization processes were observed. The non-monotonic changes in $R_2$ with increasing $x$ can be attributed to competing contributions from weak 4*f*-SOC and 3*d*-4*f* exchange. In $Tb_xFe_{1-x}$, $R_2$ increases with $x$, indicating a growing role of strong 4*f*-SOC and orbital-mediated dissipation. By contrast, $Dy_xFe_{1-x}$ and $Ho_xFe_{1-x}$ alloys exhibit decreasing $R_2$ with increasing $x$. This decrease likely stems from the weaker room-temperature magnetism of Dy and Ho compared to Gd and Tb, combined with the fact that incorporating these large RE atoms reduces the exchange coupling strength[3, 8].

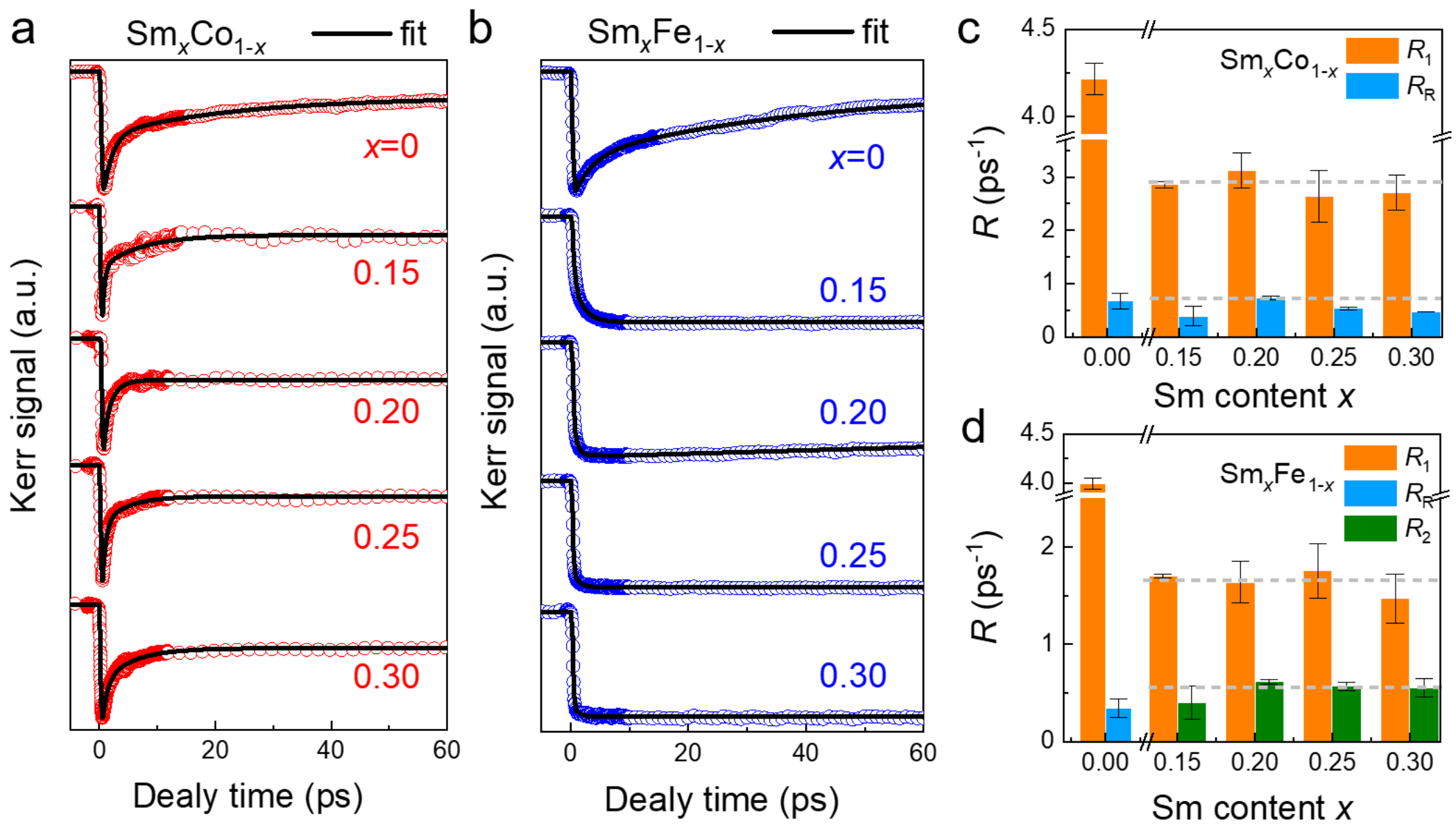

**Supplementary Figure S2. Ultrafast demagnetization dynamics in $Sm_xTM_{1-x}$ alloys.** a, b, Normalized TR-MOKE signals of $Sm_xCo_{1-x}$ films and $Sm_xFe_{1-x}$ films measured under 1.5 mJ/cm$^2$ laser excitation and 6 kOe external magnetic field, with biexponential fits overlaid. c, d, Extracted demagnetization rates $R_1$ and $R_2$ from fitting, along with recovery rate $R_R$ as functions of Sm concentration.

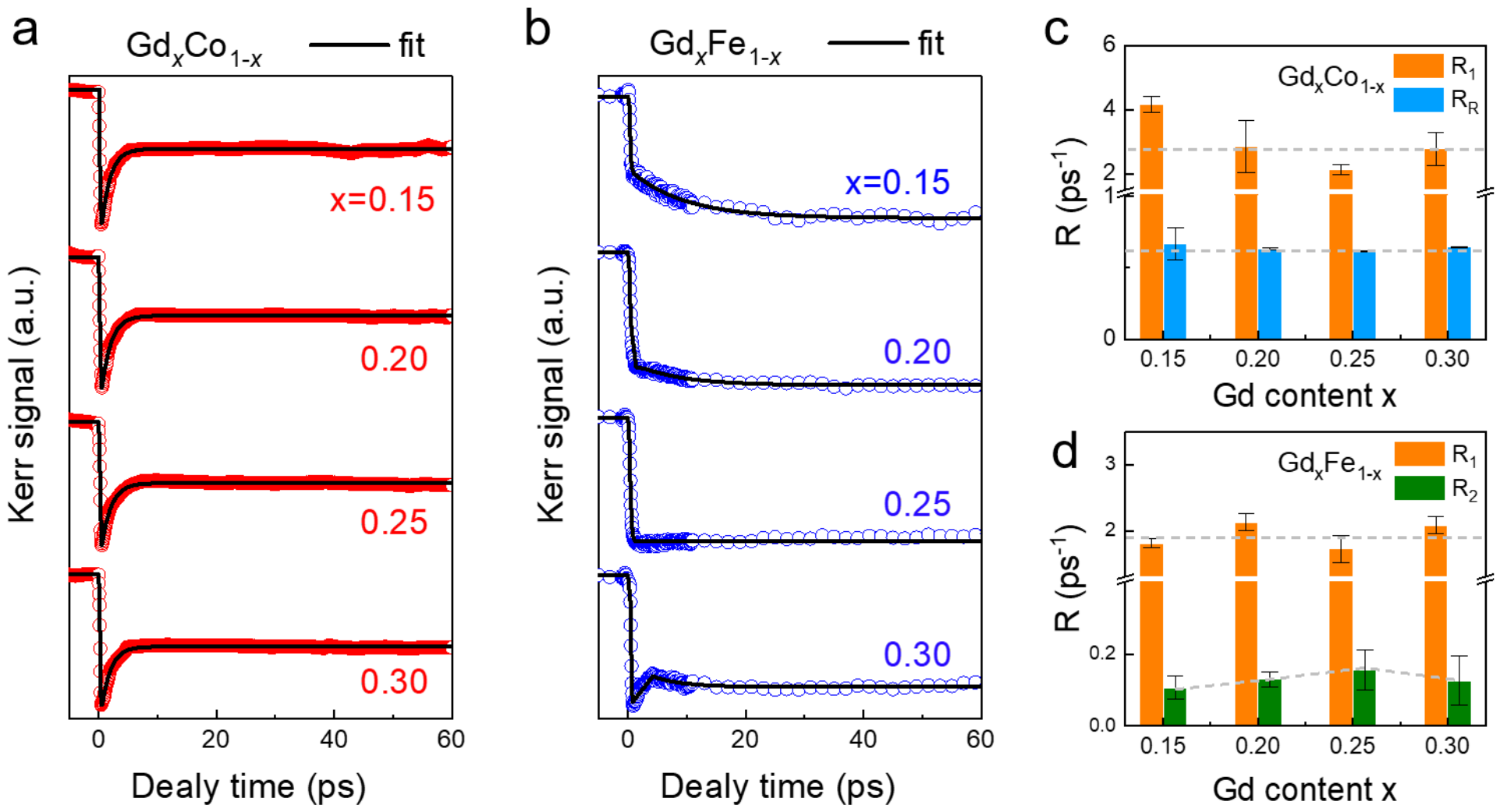

**Supplementary Figure S3. Ultrafast demagnetization dynamics in $Gd_xTM_{1-x}$ thin films.** Time-resolved Kerr signals (a, b) and extracted demagnetization rates $R_1$, $R_2$, and recovery rate $R_R$ as a function of Gd content (c, d).

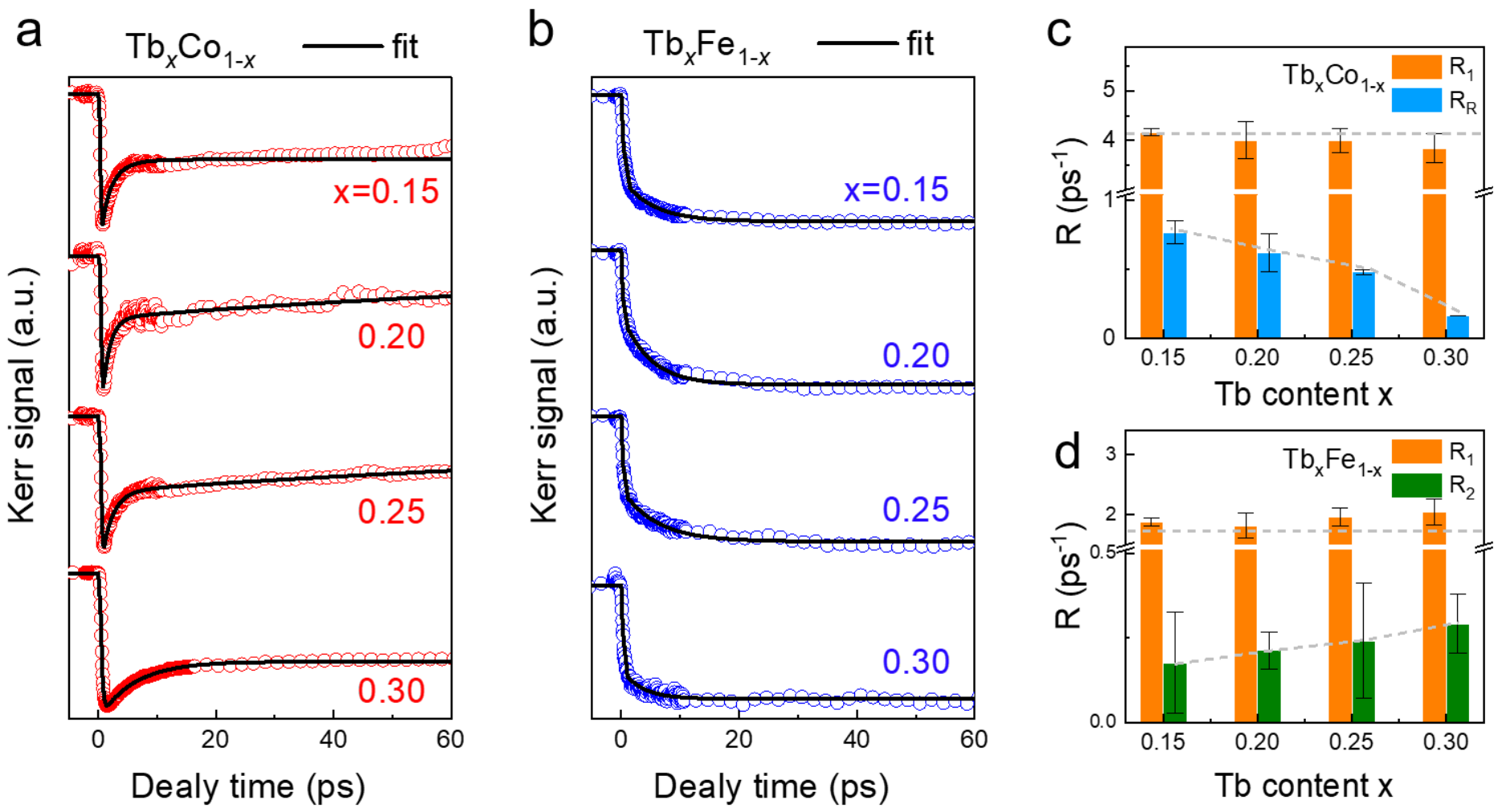

**Supplementary Figure S4. Ultrafast demagnetization dynamics in $Tb_xTM_{1-x}$ thin films.** Time-resolved Kerr signals (a, b) and extracted demagnetization rates $R_1$, $R_2$, and recovery rate $R_R$ for different Tb content (c, d).

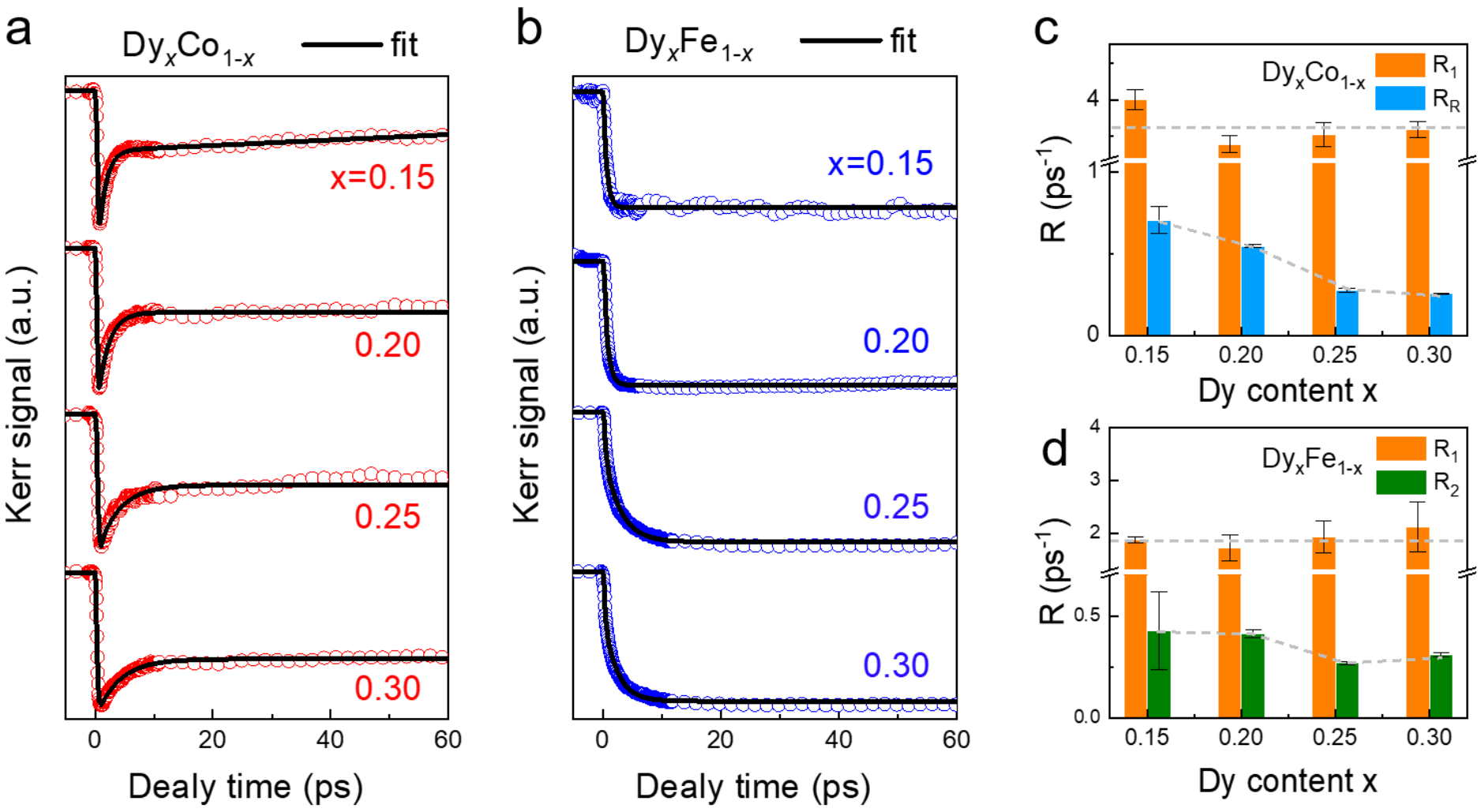

**Supplementary Figure S5. Ultrafast demagnetization dynamics in $Dy_xTM_{1-x}$ thin films.** Time-resolved Kerr signals (a, b) and extracted demagnetization rates $R_1$, $R_2$, and recovery rate $R_R$ as a function of Dy content (c, d).

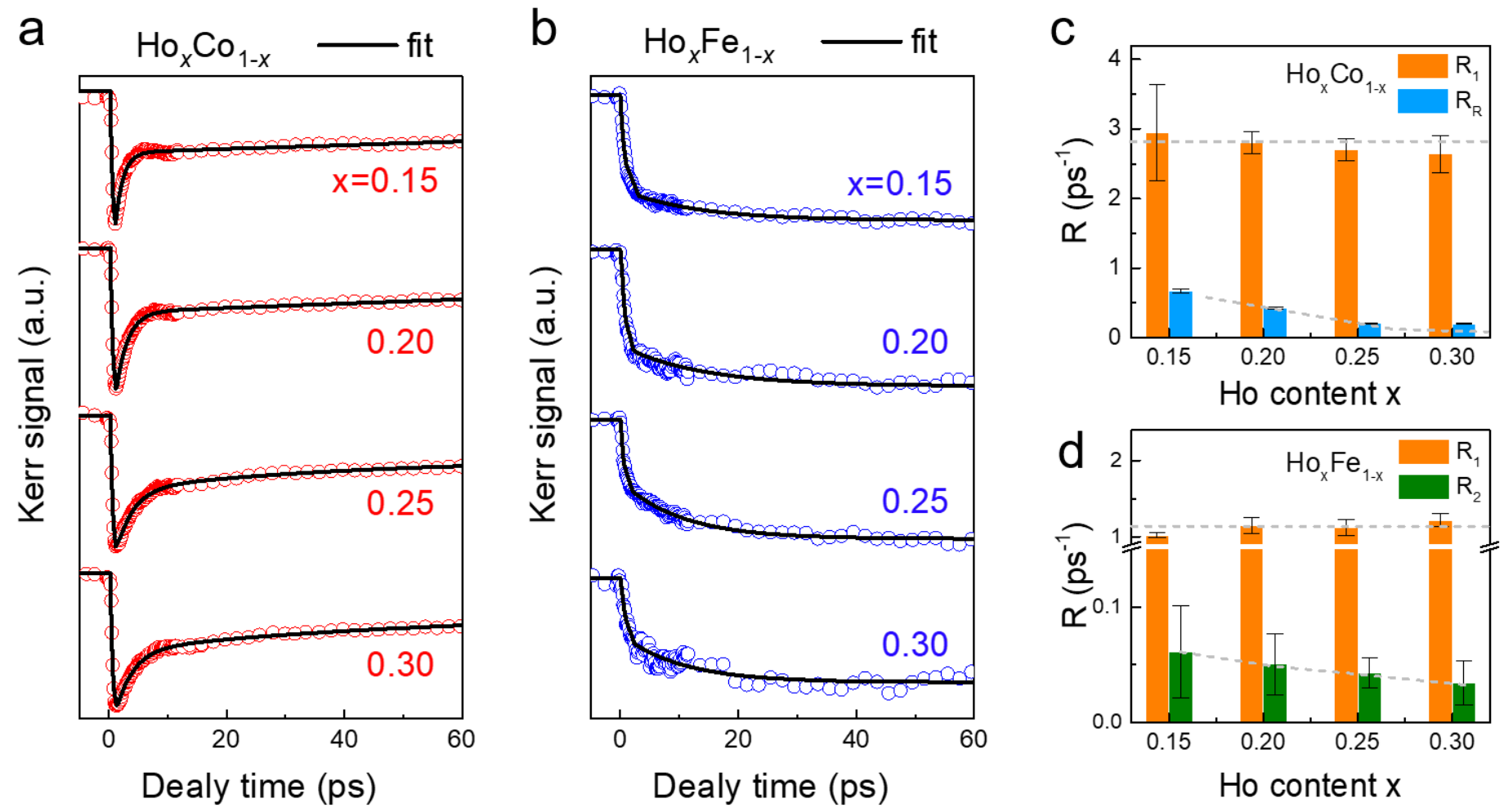

**Supplementary Figure S6. Ultrafast demagnetization dynamics in $Ho_xTM_{1-x}$ thin films.** Laser-induced magnetization dynamics (a, b) and extracted biexponential rates (c, d) for Ho-doped TM alloys.

## Supplementary Note 3: Theoretical simulations of ultrafast demagnetization by extended four temperature phenomenological model (E4TM).

Ultrafast demagnetization dynamics exhibiting single-step behavior were fitted using the following empirical function [1]

$$\theta(t) = \theta_0 + H(t)\left[B\left(1 - e^{-t/\tau_1}\right) + C\left(1 - e^{-t/\tau_R}\right)\right] \quad (1)$$

where $\theta_0$ is the initial Kerr signal and $H(t)$ is the Heaviside step function. $\tau_1$ and $\tau_R$, are the characteristic times for demagnetization and recovery, and $B$ and $C$ are the corresponding amplitudes.

For two-step demagnetization processes with continuous transitions, the dynamics are well described by [8]

$$\theta(t) = \theta_0 + H(t)\left[\Delta M_1\left(1 - e^{-t/\tau_1}\right) + \Delta M_2\left(1 - e^{-t/\tau_2}\right)\right] \quad (2)$$

Where $\tau_{1(2)}$ and $\Delta M_{1(2)}$ represent the timescales and amplitudes of the first and second demagnetization stages, respectively.

To account for spin-orbit coupling (SOC)-mediated energy transfer, we employ an extended four temperature phenomenological model (E4TM) consisting of four coupled differential equations. These describe the heat flow between four heat reservoirs, which includes electrons ($T_e$), 3*d* spins of transition metals ($T_s^{TM}$), 4*f* spins of rare earths ($T_s^{RE}$), and the lattice ($T_l$) [9, 10]:

$$C_e(T_e)\frac{dT_e}{dt} = -G_{el}(T_e - T_l) - G_{es}^{TM}(T_e - T_s^{TM}) - G_{es}^{RE}(T_e - T_s^{RE}) + P(t) \quad (3)$$

$$C_l(T_l)\frac{dT_l}{dt} = -G_{el}(T_l - T_e) - G_{soc}^{3d}(T_l - T_s^{TM}) - G_{soc}^{4f}(T_l - T_s^{RE}) \quad (4)$$

$$C_s^{TM}(T_s^{TM})\frac{dT_s^{TM}}{dt} = -G_{es}^{TM}(T_s^{TM} - T_e) - G_{soc}^{3d}(T_s^{TM} - T_l) - G_{ss}^{RE-TM}(T_s^{TM} - T_s^{RE}) \quad (5)$$

$$C_s^{RE}(T_s^{RE})\frac{dT_s^{RE}}{dt} = -G_{es}^{RE}(T_s^{RE} - T_e) - G_{soc}^{4f}(T_s^{RE} - T_l) - G_{ss}^{RE-TM}(T_s^{RE} - T_s^{TM}) \quad (6)$$

It is important to note that the spin-lattice coupling constant is explicitly denoted as $G_{SOC}$ rather than the conventional $G_{sl}$. This notation highlights the fundamental physical mechanism of the energy transfer: spin angular momentum does not directly couple to the lattice, but must first flow into the orbital degrees of freedom before transferring to the lattice. Therefore, the spin-orbit coupling (SOC) represents the primary channel and rate-limiting factor for the spin-to-lattice energy relaxation [10, 11].

Here $C_e = \gamma T_e$ is the electron specific heat with $\gamma = 714\ \mathrm{J/m^3K^2}$, and the lattice specific heat $C_l$ is assumed constant. $C_s^{TM} = 0.13 \times 10^5 \mathrm{J/m^3K}$ and $C_s^{RE} = 0.22 \times 10^5 \mathrm{J/m^3K}$ denote the spin-specific heat contributions of TM and RE sublattices, respectively. The coupling constants between electron-lattice and electron-spin are chosen as $G_{el} = 1.8 \times 10^{17} \mathrm{W/m^3K}$, $G_{es}^{TM} = 0.13 \times 10^{16} \mathrm{W/m^3K}$ and $G_{es}^{RE} = 0.6 \times 10^{15} \mathrm{W/m^3K}$. The laser heating term $P$(t) assumes a Gaussian pulse with FWHM of 100 fs and peak power density $3.5 \times 10^5 \mathrm{W/m^3}$, which is only applied to the electronic subsystem.

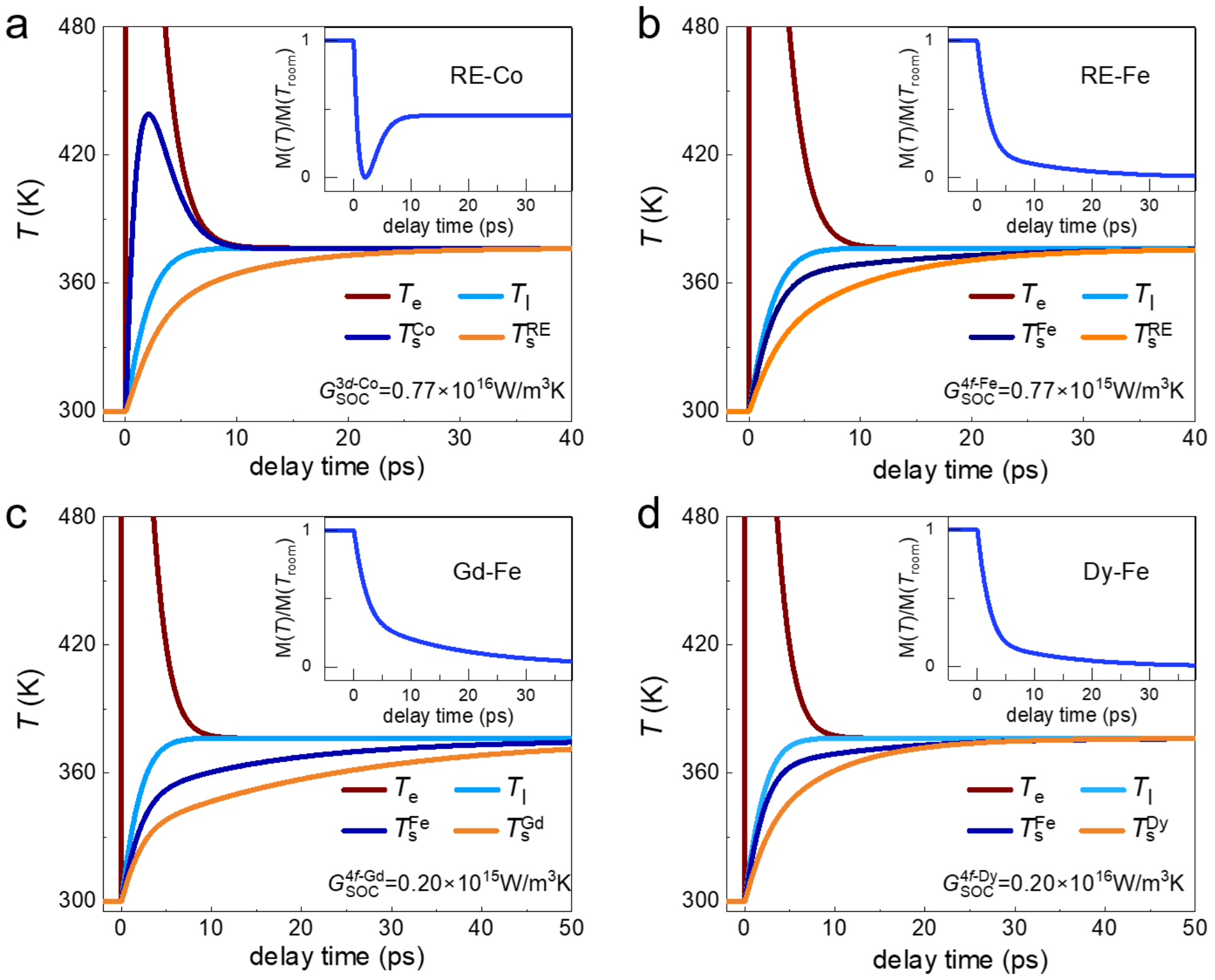

**Supplementary Figure S7. Calculation results for RE-TM alloys based on E4TM.** Temporal evolution of the heat-reservoir temperatures for electrons ($T_e$), lattice ($T_l$), TM spin subsystem ($T_s^{TM}$), and RE spin subsystem ($T_s^{RE}$) in RE-TM alloys, calculated using the extended four-temperature model. Calculation parameters are a, $G_{soc}^{3d-Co} = 0.77 \times 10^{16}\ \mathrm{W/m^3K}$, b, $G_{soc}^{3d-Fe} = 0.77 \times 10^{15}\ \mathrm{W/m^3K}$, c, $G_{soc}^{4f-Gd} = 0.20 \times 10^{15}\ \mathrm{W/m^3K}$, and d, $G_{soc}^{4f-Dy} = 0.20 \times 10^{16}\ \mathrm{W/m^3K}$. Insets show the corresponding magnetization dynamics derived from spin temperature evolution using the Bloch $T^{3/2}$ law.

Numerical integration of Eqs. (3) - (6) using the Runge-Kutta method reproduces key experimental features. In RE-Co alloys (Fig. S7a), we set $G_{\mathrm{soc}}^{3d\text{-Co}} = 0.77 \times 10^{16}$ W/m$^3$K, which is one order of magnitude greater than the intersublattice coupling constant $G_{\mathrm{ss}}^{\text{RE-Co}} = 0.5 \times 10^{15}$W/m$^3$K [9, 10]. The Co spin temperature $T_{\mathrm{s}}^{\mathrm{Co}}$ increases rapidly to ~440 K within 2 ps, accompanied by a synchronous increase in $T_{\mathrm{l}}$, indicating efficient energy transfer from the spin subsystem to the lattice via orbital-lattice coupling. In the RE-Fe system (Fig. S7b), we adopt $G_{\mathrm{soc}}^{3d\text{-Fe}} = 0.77 \times 10^{15}$ W/m$^3$K and $G_{\mathrm{ss}}^{\text{RE-Fe}} = 0.3 \times 10^{15}$ W/m$^3$K. Under femtosecond laser excitation, $T_{\mathrm{s}}^{\mathrm{Fe}}$ rises to ~350 K at 2.8 ps, while $T_{\mathrm{l}}$ increases more gradually, reaching ~ 375 K at 5 ps. Notably, after ~ 2 ps, $T_{\mathrm{l}}$ surpasses $T_{\mathrm{s}}^{\mathrm{Fe}}$ and continues to lead until thermal equilibrium is established at around 25 ps. This delayed spin-lattice equilibration reflects the relatively weak SOC in Fe, which hinders efficient angular momentum transfer from the spin subsystem to the lattice. In this scenario, intersublattice 3*d*-4*f* exchange coupling becomes significant. Energy is initially transferred from the Fe 3*d* spins to the RE 4*f* spins, some of which is subsequently dissipated into the lattice via RE-SOC-mediated orbital-lattice interactions, while the remainder flows back into the 3*d*-spin reservoir, effectively reheating it and slowing down the overall demagnetization process.

Beyond the dominant role of 3*d*-SOC in RE-Co systems, the contribution of 4*f*-SOC becomes particularly significant in RE-Fe alloys. To illustrate this effect, we performed comparative calculations on two representative systems: GdFe and DyFe, which differ markedly in their orbital angular momentum quantum numbers ($Q_{\mathrm{L}}$= 0 for Gd and $Q_{\mathrm{L}}$=5 for Dy). In the simulations, we set $G_{\mathrm{soc}}^{4f\text{-Gd}} = 0.20 \times 10^{15}$ W/m$^3$K and $G_{\mathrm{soc}}^{4f\text{-Dy}} = 2.0 \times 10^{15}$ W/m$^3$K, as shown in Figs S7c-S7d. In the GdFe system (Fig. S7c), both $T_{\mathrm{s}}^{\mathrm{Fe}}$ and $T_{\mathrm{s}}^{\mathrm{Gd}}$ increase slowly and only reach thermal equilibrium only after ~50 ps. This prolonged timescale is attributed to the weak SOC in Gd, which limits efficient energy dissipation. Consequently, energy transferred from the Fe 3*d* spins to the Gd 4*f* spins via 3*d*-4*f* coupling is largely reflected back into the 3*d* subsystem, causing a re-heating effect, as reported in earlier studies[15]. In contrast, the DyFe system (Fig. S7d) exhibits markedly faster spin dynamics. Both $T_s^{Fe}$ and $T_s^{Dy}$ increase more rapidly and reach thermal equilibrium within ~20 ps. This acceleration is attributed to Dy's strong 4*f*-SOC, which facilitates efficient orbital-lattice energy transfer. Although part of the energy still flows back from the Dy 4*f* spins to the Fe 3*d* spins via inter-sublattice coupling, a substantial portion is dissipated directly into the lattice through Dy's 4*f* orbital channel. This additional dissipation route mitigates energy backflow to the 3*d* subsystem and accelerates overall relaxation.

**Supplementary Note 4: Ultrafast demagnetization dynamics in Ni-doped RE-Co alloys.**

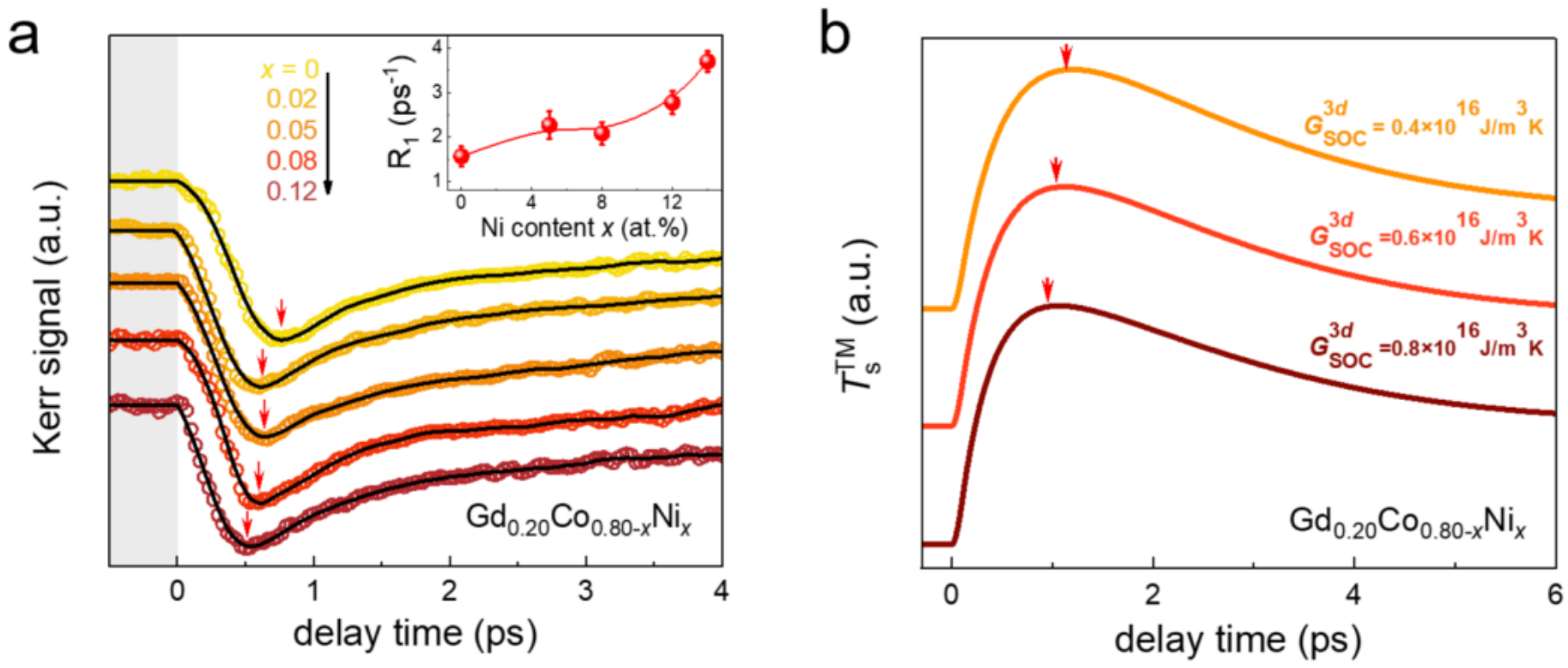

**Supplementary Figure S8. Ultrafast demagnetization dynamics of Ni-doped RE-Co systems.** a, TR-MOKE measurements of ultrafast magnetization dynamics in Ni-doped $Gd_{0.20}Co_{0.80}$ thin films under a pump fluence of 1.5 mJ/cm$^2$ and an applied magnetic field of $H_{ext}$ = 6 kOe. Red arrows indicate the shift of the TR-MOKE signal extrema toward shorter pump-probe delays with increasing Ni content. Solid black lines represent biexponential fits to extract the first-step demagnetization time constant ($\tau_1$). Insets: Dependence of demagnetization rate $R_1$ on Ni content for alloy system. b, Simulated temporal evolution of the TM spin temperature $T_s^{TM}$ based on the extended four-temperature model for $Gd_{0.20}Co_{0.80-x}Ni_x$ alloys. Calculations are performed using $G_{soc}^{3d} = 0.4 \times 10^{16}\ \mathrm{W/m^3K}$, $G_{soc}^{3d} = 0.6 \times 10^{16}\ \mathrm{W/m^3K}$, and $G_{soc}^{3d} = 0.8 \times 10^{16}\ \mathrm{W/m^3K}$, corresponding to different Ni doping levels. The 3*d*-4*f* spins coupling constants were set to $G_{ss}^{Gd\text{-}TM} = 0.5 \times 10^{15}\mathrm{W/m^3K}$ for $Gd_{0.20}Co_{0.80-x}Ni_x$.

To further validate the universality of SOC-modulated ultrafast demagnetization in RE-TM systems, we also performed controlled doping of Ni (SOC strength ≈ 110 meV) into GdCo alloys. Our comparative analysis across RE-Fe and RE-Co systems revealed a clear positive correlation between 3*d*-SOC strength and the demagnetization rate ($R$). Building on this insight, we systematically increased the 3*d*-SOC strength in GdCo alloys by substituting small amounts of Co with Ni, while preserving the overall crystal structure and magnetic ordering. According to our theoretical framework, the demagnetization rate should increase monotonically with Ni content, due to Ni's inherently stronger SOC compared to Co. The enhanced SOC facilitates more efficient AM transfer from the 3*d* spin subsystem to the lattice by improving the coupling strength of the 3*d* orbital-lattice coupling channel. As shown in Figs. S8a, GdCoNi alloys display one-step, sub-ps demagnetization followed by rapid recovery, mirroring the dynamics of undoped RE-Co alloys, signifying that the ultrafast demagnetization process remains dominated by the strong SOC inherent in the CoNi matrix.

To quantitatively analyze the impact of Ni doping on ultrafast demagnetization, we fitted the transient TR-MOKE signals using a biexponential function, as presented in the insets of Fig. S8a. The extracted first-step demagnetization rate $R_1$ clearly increases with rising Ni concentration in both alloy systems. This trend is in agreement with prior studies [13, 14], which reported that incorporating elements with higher SOC leads to faster demagnetization. These findings further confirm that tuning of 3*d*-SOC strength through compositional engineering such as Ni doping, can substantially improve the efficiency of energy dissipation mediated by 3*d*-electron orbitals in RE-Co alloys. Furthermore, the experimentally observed increase in the demagnetization rate $R_1$ is successfully reproduced by the E4TM with enhanced 3*d*-SOC. In the numerical calculations, all parameters were held constant as in the original calculations, except for $G_{\text{soc}}^{3d}$. As the Ni concentration increases, the effective 3*d*-SOC strength is enhanced, leading to faster energy transfer from the spin subsystem to the lattice. The simulated temporal profiles of the TM spin temperature $T_{\text{s}}^{\text{TM}}$ for $Gd_{0.20}Co_{0.80-x}Ni_x$ (Fig. S8b) clearly shows this acceleration in spin-lattice energy transfer dynamics. The simulations were conducted with $G_{\text{soc}}^{3d}$ systematically varied from $0.4\times10^{16}$ J/m$^3$K to $0.8\times10^{16}$ J/m$^3$K. A pronounced increase in $R_1$ with increasing $G_{\text{soc}}^{3d}$ was observed, in good qualitative agreement with experimental trends. It is worth noting that, due to the phenomenological nature of the E4TM, relatively large changes in $G_{\text{soc}}^{3d}$ are required to yield distinguishable differences in the simulated dynamics. Nevertheless, the model effectively captures the essential experimental behavior.

**Supplementary Note 5: THz emission of $Gd_xFe_{1-x}$ and $Gd_xCo_{1-x}$.**

Terahertz (THz) emission measurements are performed on Sapphire/Ti(3nm)/ $Gd_xFe_{1-x}$(20nm)/Al(3nm) and Sapphire/Ti(3nm)/$Gd_xCo_{1-x}$(20nm)/Al(3nm) thin films using a THz time-domain spectrometer with an in-plane saturating magnetic field. Representative THz emission spectra near the compensation composition are shown in Fig. S9a. For $Gd_{0.2}Co_{0.8}$, a significantly stronger THz transient is observed compared to $Gd_{0.15}Fe_{0.85}$, indicating that the transient orbital current, proportional to the emitted THz field, is substantially larger in the Co-based alloy. This observation suggests that Co 3*d* orbitals mediate more efficient energy transfer from the spin system to the lattice in RE-TM ferrimagnet.

Figure S9b presents the Gd composition dependence of the THz emission intensity for both alloy systems. For $Gd_xCo_{1-x}$, the THz intensity increases with Gd content up to $x \approx 0.2$ and then decreases. This nonmonotonic behavior correlates with the composition-driven reorientation of the magnetic moment from in-plane to out-of-plane, which modifies the efficiency of ultrafast spin-to-charge conversion. A similar trend is observed in $Gd_xFe_{1-x}$, with a peak near $x \approx 0.15$. Notably, the THz emission intensities of pure Co and Fe reference films are comparable, consistent with their similar demagnetization time scales. This likely originates from Co's stronger net spin magnetic moment (vs. Fe), which enhances its propensity for orbital hybridization with RE elements. The underlying physics will be explored in future studies.

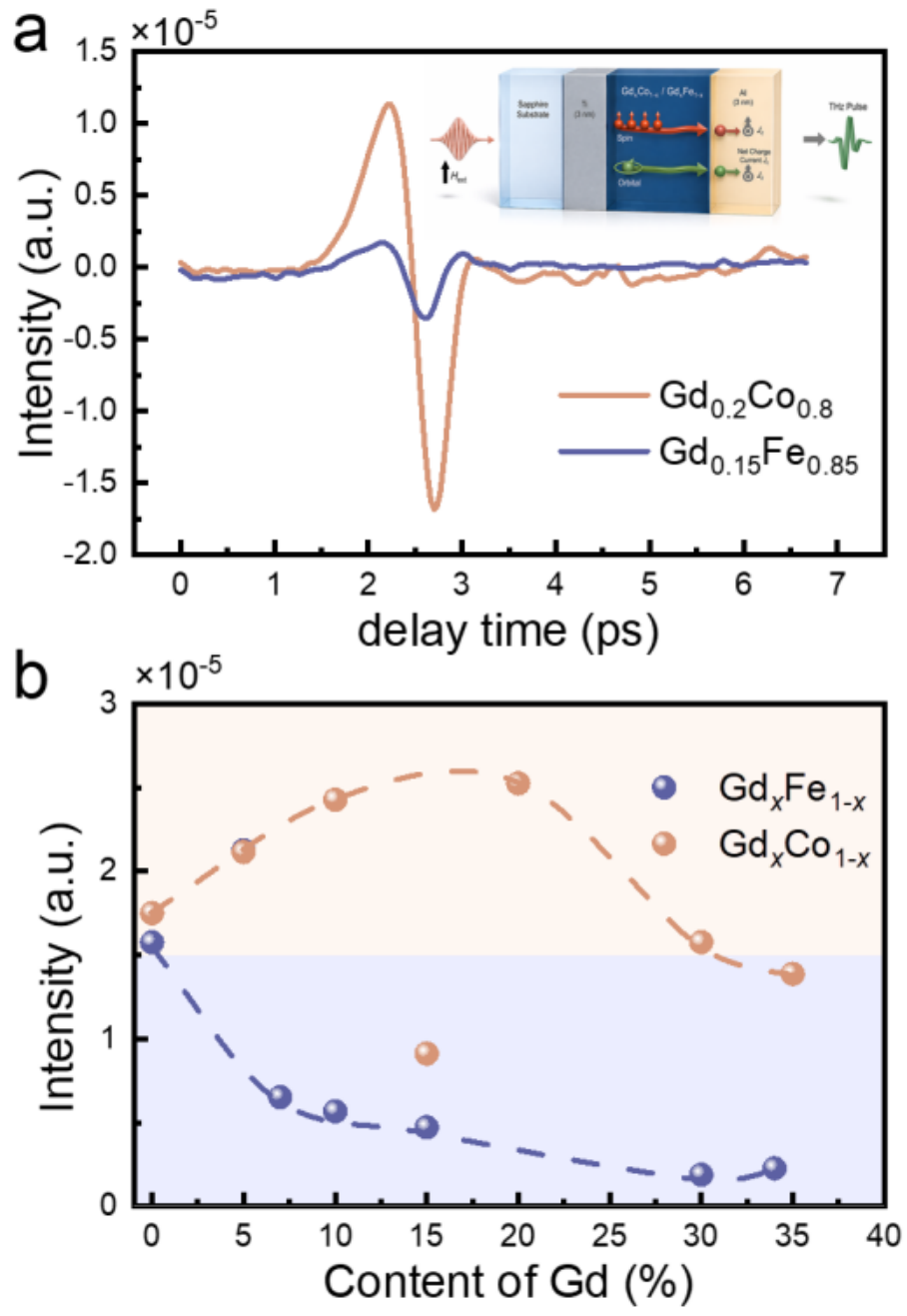

**Supplementary Figure S9. THz emission of $Gd_xFe_{1-x}$ and $Gd_xCo_{1-x}$.** a, Typical THz emission spectra of Ti(3nm)/$Gd_{0.2}Co_{0.8}$(20nm)/Al(3nm) and Ti(3nm)/ $Gd_{0.15}Fe_{0.85}$(20nm)/Al(3nm) thin films. b, THz emission intensity of Ti(3nm)/$Gd_xFe_{1-x}$(20nm)/Al(3nm) and Ti(3nm)/$Gd_xCo_{1-x}$(20nm)/Al(3nm) alloys as a function of Gd composition.

The markedly higher THz emission efficiency in GdCo compared to GdFe across all compositions highlights the distinct roles of Co and Fe 3*d* orbitals in RE-TM alloys. Several factors may contribute to this difference. First, the stronger spin-orbit coupling in Co compared to Fe enhances the efficiency of ultrafast spin-to-orbital current conversion, which is directly probed by THz emission. Second, Co 3*d* orbitals exhibit stronger hybridization with Gd 5*d* states, facilitating more efficient interfacial spin transport and modifying the electronic structure near the Fermi level[17]. This hybridization effect enables faster angular momentum transfer from the photoexcited RE sublattice to the TM sublattice [18]. Third, within the framework of superdiffusive spin transport, the spin-dependent mean free paths and scattering rates differ significantly between Co and Fe, which might influence the net spin current generated upon laser excitation [16, 19]. The extreme spin asymmetry in Co (majority-spin mean free path ≈55 Å, minority-spin ≈6 Å) leads to highly efficient spin-to-orbital conversion at the Gd/Co interface, whereas the more symmetric and opposite spin-dependent transport in Fe yields a markedly weaker orbital current. This disparity is further reinforced by the distinct 3*d*-5*d* hybridization characteristics of the two transition metals. The *d*-band filling model indicates that when the 3*d* band is nearly full, positive exchange interactions dominate, whereas a half-filled band leads to competition between negative and positive interactions [20]. Consequently, Co ($3d^8$) exhibits stronger ferromagnetic coupling with Gd than Fe ($3d^6$), which is closer to the half-filled configuration. Although Co and Fe exhibit similar ultrafast demagnetization times in their elemental forms, their distinct electronic structures lead to qualitatively different spin dynamics in RE-TM heterostructures.

## Supplementary References